# Proton-Irradiation-Immune Electronics Implemented with Two-Dimensional Charge-Density-Wave Devices

A. Geremew[1], F. Kargar[1], E. X. Zhang[2], S. E. Zhao[2], E. Aytan[1], M. A. Bloodgood[3], T. T. Salguero[3], S. Rumyantsev[1,4], A. Fedoseyev[5], D. M. Fleetwood[2] and A. A. Balandin[1,♣]

[1]Nano-Device Laboratory, Department of Electrical and Computer Engineering, Materials Science and Engineering Program, University of California, Riverside, California 92521 USA

[2]Department of Electrical Engineering and Computer Science, Vanderbilt University, Nashville, Tennessee 37235 USA

[3]Department of Chemistry, University of Georgia, Athens, Georgia 30602 USA

[4]Center for Terahertz Research and Applications, Institute of High-Pressure Physics, Polish Academy of Sciences, Warsaw, 01-142 Poland

[5]Ultra Quantum Inc., Huntsville, Alabama 35758 USA

**Proton radiation damage is an important failure mechanism for electronic devices in near-Earth orbits, deep space and high energy physics facilities [1–4]. Protons can cause ionizing damage and atomic displacements, resulting in device degradation and malfunction [5–10]. Shielding of electronics increases the weight and cost of the systems but does not eliminate destructive single events produced by energetic protons [8,10]. Modern electronics based on semiconductors – even those specially designed for radiation hardness – remain highly susceptible to proton damage. Here we demonstrate that room temperature (RT) charge-density-wave (CDW) devices with quasi-two-dimensional (2D) 1T-TaS$_2$ channels show *remarkable* immunity to bombardment with 1.8 MeV protons to a fluence of at least $10^{14}$ H$^+$cm$^{-2}$. Current-**

---

[♣] Corresponding author (AAB): balandin@ece.ucr.edu ; web: https://balandingroup.ucr.edu/

A. Geremew, F. Kargar, E. X. Zhang, S. E. Zhao, E. Aytan, M. A. Bloodgood, T. T. Salguero, S. Rumyantsev, A. Fedoseyev, D. M. Fleetwood and A. A. Balandin (Dec 2018)

**voltage *I-V* characteristics of these 2D CDW devices do not change as a result of proton irradiation, in striking contrast to most conventional semiconductor devices or other 2D devices. Only negligible changes are found in the low-frequency noise spectra. The radiation immunity of these "all-metallic" CDW devices can be attributed to their two-terminal design, quasi-2D nature of the active channel, and high concentration of charge carriers in the utilized CDW phases. Such devices, capable of operating over a wide temperature range, can constitute a crucial segment of future electronics for space, particle accelerator and other radiation environments.**

A. Geremew, F. Kargar, E. X. Zhang, S. E. Zhao, E. Aytan, M. A. Bloodgood, T. T. Salguero, S. Rumyantsev, A. Fedoseyev, D. M. Fleetwood and A. A. Balandin (Dec 2018)

The future of human and unmanned space exploration depends crucially on the development of new electronic technologies that are immune to space radiation, which consists primarily of protons, electrons, and cosmic rays [1–4]. The penetrating energetic radiation of deep space produces negative impacts on not only biological entities but also the electronic systems of space vehicles. Electronics capable of operating in high-radiation environments are also needed for monitoring nuclear materials, medical diagnostics, radiation treatments, nuclear reactors and particle accelerators [5–12]. Shielding of electronic systems in space is limited to lower-energy electrons and protons. High-energy proton irradiation causes ionizing damage by generating excess charges at the interface regions in the complementary metal-oxide-semiconductor (CMOS) transistors and other typical microelectronic devices and integrated circuits [8–14]. This type of damage results in the changes in the threshold voltages and source–drain currents, potentially leading to device or system failure. Protons also can induce displacement, which typically leads to the formation of point defects in semiconductors. These are electronic trapping states that often reveal themselves by increases in low-frequency noise (LFN) [15,16] Noise increases beyond system tolerance limits is therefore an additional challenge to electronics in high-radiation environments. Shielding, the use of conventional radiation-hardened technologies and backup devices, increases system complexity and costs while not providing complete protection from proton-induced damage. This situation motivates the exploration of new materials and innovative device designs that are immune to proton irradiation.

A promising approach for developing radiation-immune electronics is the use of CDW effects, which appear in some metallic systems [17]. For many years, classical CDW materials like the bulk quasi-one-dimensional (1D) metals $NbSe_3$ and $TaSe_3$ were considered interesting from the physics point of view but impractical owing to the low temperatures of the CDW phase transitions. Recently, however, their status has changed. New studies have demonstrated that transitions between the incommensurate (IC) and

A. Geremew, F. Kargar, E. X. Zhang, S. E. Zhao, E. Aytan, M. A. Bloodgood, T. T. Salguero, S. Rumyantsev, A. Fedoseyev, D. M. Fleetwood and A. A. Balandin (Dec 2018) nearly commensurate (NC) CDW states in some quasi-2D materials, which happen above RT, can be triggered effectively by electric field [18]. Specifically, 1$T$-TaS$_2$ shows a transition from the NC-CDW to IC-CDW phase at 350 K. An electric field applied along the device channel can trigger the NC-to-IC phase transition, accompanied by an abrupt change in the electrical conductivity and hysteresis loop, at *various* temperatures [19–21]. Additional work has shown that such CDW current-voltage, *I-V*, characteristics can be utilized for implementing voltage controlled oscillators (VCO), modulators and logic gates [18,22,23]. These studies also suggest that CDW devices can be low-power [22,23] and extremely fast—up to terahertz frequencies [18].

Metals are inherently more radiation tolerant than semiconductors owing to much higher carrier densities, *i.e.* $n_e \geq 10^{20}$ cm$^{-3}$ in metals *vs.* $10^{10}$ cm$^{-3}$ – $10^{19}$ cm$^{-3}$ in semiconductors. One can also expect that the quasi-2D geometry of the channels and two-terminal structure of CDW devices will make them less susceptible to radiation. There is a possibility that protons will pass through the thin-film channel without inflicting significant damage. Thus, the *most important* test for the radiation hardness of quasi-2D CDW devices is investigation of their susceptibility to *proton irradiation*. In this study, we show that CDW devices with quasi-2D 1$T$-TaS$_2$ channels, operating at RT, reveal *high immunity* to 1.8 MeV proton radiation of up to a fluence of at least $10^{14}$ P$^+$/cm$^2$. This discovery demonstrates the promise of CDW devices for use in critical electronic systems for space exploration and other radiation environments.

For this study, high-quality 1$T$-TaS$_2$ crystals were prepared by the chemical vapor transport method (see *Supplemental Materials*). To avoid air oxidation and chemical contamination, devices with 1$T$-TaS$_2$ channels and Ti/Au contacts were fabricated using the shadow mask method (see *Methods*). To investigate the intrinsic radiation hardness, we intentionally did not use a hexagonal boron nitride channel capping, a typical design approach for quasi-2D devices. The schematic, layered structure, typical *I-V*

A. Geremew, F. Kargar, E. X. Zhang, S. E. Zhao, E. Aytan, M. A. Bloodgood, T. T. Salguero, S. Rumyantsev, A. Fedoseyev, D. M. Fleetwood and A. A. Balandin (Dec 2018) characterization with load resistance, and oscillations resulting from a properly biased 1$T$-TaS$_2$ device are shown in Figure 1a. An optical microscopy image of an array of 1$T$-TaS$_2$ devices with different CDW channel lengths is presented in Figure 1b.

RT proton irradiation was performed under vacuum conditions using the Vanderbilt Pelletron [24]. To allow for better comparison with the prior reports on other devices, we selected the proton energy of 1.8 MeV. Experiments were performed at fluences of $1\times10^{12}$ cm$^{-2}$, $3\times10^{12}$ cm$^{-2}$, $1\times10^{13}$ cm$^{-2}$, $3\times10^{13}$ cm$^{-2}$, and $1\times10^{14}$ cm$^{-2}$. The flux for the proton irradiation was around $5\times10^{10}$ protons (s$^{-1}\cdot$cm$^{-2}$). The investigated irradiation dose in the present study (up to $10^{14}$ P$^+$/cm$^2$) is higher than would be observed in any realistic space environment, and comparable to that experienced in nuclear reactor loss-of-coolant accidents or future high-luminosity particle accelerator applications. Details of the proton testing are provided in the *Methods*. The *I-V* characteristics were checked after each irradiation step for several devices. After the final irradiation dose, LFN characteristics were measured. Raman spectroscopy and transmission electron microscopy were used to assess possible damage to 1$T$-TaS$_2$ channels. Figure 1c shows Raman spectra of 1$T$-TaS$_2$ before and after proton irradiation to a total fluence of $10^{14}$ cm$^{-2}$. The spectra did not show any signs of the crystal lattice damage. In Figure 1d, we present *I-V* characteristics of the representative quasi-2D 1$T$-TaS$_2$ CDW device before irradiation and after different proton radiation doses.

[Figure 1]

The measured *I-V* characteristics clearly show the linear low-field region followed by the hysteresis due to the NC-CDW to IC-CDW transition, consistent with all prior reports on such devices [18–23]. The hysteresis region is the *functional* part of *I-V* curves, essential for operation of the CDW VCOs and logic gates [18,22,23] (see Figure 1a). For comparison with other materials and technologies, it is convenient to analyze the relative changes in the

A. Geremew, F. Kargar, E. X. Zhang, S. E. Zhao, E. Aytan, M. A. Bloodgood, T. T. Salguero, S. Rumyantsev, A. Fedoseyev, D. M. Fleetwood and A. A. Balandin (Dec 2018) current, $\Delta I/I$, and voltage, $\Delta V/V$, due to irradiation. In conventional devices, $I$ is the source-drain current, while $V$ is the threshold, $V_T$, or drain, $V_D$, voltage. In our case, $V$ is the onset of the resistive switching, which is a close analog of $V_T$, and $I$ is the corresponding current (see the limiting values in the hysteresis loop). Naturally, $\Delta I/I \approx 1$ indicates an extremely high radiation damage while $\Delta I/I \approx 0$ means no damage. Analyzing Figure 1d one can sees that quasi-2D CDW devices do not show any noticeable proton radiation damage, even after a total fluence of $10^{14}$ cm$^{-2}$ of 1.8 MeV protons. The functional hysteresis window preserves its shape after all irradiation steps, and changes in the relevant *I-V* characteristics were as low as $\Delta I/I < 10^{-5}$ and $\Delta V/V < 1.4 \times 10^{-5}$. These values are *extremely* small compared to all other 2D devices reported in the literature. Table I provides a comparison of 1$T$-TaS$_2$ CDW devices to the best devices implemented with other 2D materials and tested for proton irradiation damage [25–29]. The total ionizing dose (TID) and the charge carrier concentration, $n$, are either taken directly from the reports or calculated from the data provided. One should note that all other 2D devices use a field-effect transistor (FET) concept, which make them more susceptible to proton and x-ray damage due to charge trapping in surrounding insulators or at the device-insulator interfaces. Additional reference information for radiation damage, including x-rays and gamma-rays, for all types of devices are provided in the *Supplemental Materials*.

**Table I: Proton bombardment effects on quasi-2D devices**

| 2D Devices | $n$ (cm$^{-2}$) | $\Delta I/I$ | $\Delta V/V$ | TID rad (SiO$_2$) | Energy (MeV) | Ref |
|---|---|---|---|---|---|---|
| 1$T$-TaS$_2$ | $10^{18}$ | $3 \times 10^{-5}$ | $1.4 \times 10^{-5}$ | $2.0 \times 10^8$ | 1.80 | This work |
| Graphene | $10^{13}$ | 1.00 | 0.99 | $4.0 \times 10^9$ | 5.00 | 25 |
| Graphene | $10^{13}$ | 0.33 | - | $4.0 \times 10^8$ | 15.0 | 26 |
| Graphene | $10^{13}$ | 0.67 | 0.99 | $2.0 \times 10^{10}$ | 5.00 | 27 |
| MoS$_2$ | $10^{12} - 10^{13}$ | 1.00 | 1.00 | $2.0 \times 10^8$ | 10.0 | 28 |
| WSe$_2$ | $10^{12} - 10^{13}$ | 0.11 | 0.99 | $2.0 \times 10^8$ | 2.00 | 29 |

The negligible changes in $\Delta I/I$ and $\Delta V/V$ indicate insignificant total ionizing damage. To examine other possible negative effects of proton bombardment, *e.g.* displacement damage, we conducted LFN measurements, following standard protocols (see *Methods*).

A. Geremew, F. Kargar, E. X. Zhang, S. E. Zhao, E. Aytan, M. A. Bloodgood, T. T. Salguero, S. Rumyantsev, A. Fedoseyev, D. M. Fleetwood and A. A. Balandin (Dec 2018)

The high energy – high fluence proton irradiation of the device channel can result in physical displacement of atoms leading to creation of defects. The higher density of defects, with various time constants, often results in increased LFN, irrespective of its mechanism, *i.e.* carrier number *vs.* mobility fluctuations [15,16]. This relationship emphasizes the importance of LFN spectroscopy for assessing the radiation damage related to defect creation [10]. Figure 2a shows the voltage-referred noise power spectral density, $S_V$, as a function of frequency, $f$, for several voltage biases, $V_D$, before proton irradiation and after the final irradiation step ($10^{14}$ P$^+$/cm$^2$ bombardment). All noise spectra are of the $1/f$ type without any traces of the generation – recombination bulges. The most important observation is that the noise level hardly changes after high-dose proton irradiation. The current noise spectral density, $S_I$, is presented as a function of current, $I$, in Figure 2b. It is known from the noise theory that $S_I$ proportionality to $I^2$ is indicative of conventional $1/f$ noise, without signs of Joule heating, electromigration or material degradation. Figures 2c and 2d show the *I-V* characteristics and the normalized noise spectral density, $S_I/I^2$ ($f = 10$ Hz) of the quasi-2D 1*T*-TaS$_2$ device after proton bombardment, as a function of the applied voltage, $V_D$. The inset to Figure 2c shows the low-bias resistivity as a function of temperature across the NC to IC CDW phase transition. The overall LFN level is rather low, $S_I/I^2 \sim 10^{-12}$ Hz$^{-1}$ – $10^{-10}$ Hz$^{-1}$, compared to mature conventional technologies [15], and reaches a maximum at $V_D$=1.4 V. A comparison of Figures 2c and 2d indicates that the noise peak is associated with the IC to NC CDW phase transition, which is demonstrated here to be unaffected by proton bombardment. LFN attains its maxima at phase transitions, which explains its significance as a spectroscopy tool for CDW effects [30].

[Figure 2]

To understand better the proton bombardment immunity of quasi-2D 1*T*-TaS$_2$ CDW devices, we performed simulation of the impact of proton radiation using the stopping and range of ions in solids (SRIM) software tools [31]. The SRIM method is based on the Monte Carlo approach with the binary collision approximation and random selection of the impact parameters of the colliding ions [31,32]. The simulated device structure consisted of the layers of 1$T$-TaS$_2$ (90 nm), SiO$_2$ (300 nm) and Si (10 μm), subjected to a 1.8 MeV proton beam. The simulation results suggest that the damage caused by atom displacements is quite low, at the level of 10$^{-6}$ per ion. This means that the long-term device characteristics are not impaired even under high proton irradiation. Protons primarily pass through the 1$T$-TaS$_2$ device channel without causing damage and deposit their energy in the Si substrate. The average energy lost to recoils, on the order of 1 meV, is insignificant compared to the initial proton energy of 1.8 MeV. The simulation data are presented in the *Supplementary Materials*.

Considering the data in Table I and the results of SRIM simulations, one can attribute the proton radiation immunity of CDW devices to the thin-film structure of the active channel, the two-terminal design, and the high concentration of the charge carriers in metallic 1$T$-TaS$_2$. The protons penetrate the thin film channel without causing significant damage and deposit most of their energy in the depth of Si substrate, without affecting the device operation. The same process happens in quasi-2D FETs [26–29,33,34]. However, since graphene and MoS$_2$ FETs rely on gate bias for their operation, even small charge accumulation in the surrounding dielectric layers or at the interface leads to substantial changes in $\Delta I/I$ and $\Delta V/V$. Similar considerations apply to carbon nanotube FETs, with the best radiation hardened devices attaining $\Delta I/I \approx 0.06$ and $\Delta V/V \approx 0.08$ under 40-keV X-ray irradiation to 2×10$^6$ rad/SiO$_2$ [35], a change more than two orders of magnitude larger than that experienced by devices in this work. The quasi-2D CDW 1$T$-TaS$_2$ devices are two-terminal, operated by changing the in-plane voltage bias. The load resistor is connected in series, and it is made of inherently radiation-tolerant metal. The absence of the traditional gate and gate dielectric removes components susceptible to radiation damage mechanisms related to dielectric or surface charge accumulation. The high concentration of carriers in 1*T*-TaS$_2$, both in NC and IC CDW states, which is comparable to metals, further reduces possible fluctuations in *ΔI/I* and *ΔV/V* owing to the defect creation and electron trapping. In comparison to other two-terminal resistive switching devices, the CDW devices perform better due to their quasi-2D thin-film channel design and substantially higher carrier densities. The radiation hardness of conventional, well-established devices technologies, such as Si CMOS, GaN and SiC FETs varies considerably depending on the device design, dimensions and special radiation hardness provisions. Typical *ΔV/V* values for the conventional technologies are in the range 0.005-0.99 [6–14], which are substantially larger than those observed for the quasi-2D 1*T*-TaS$_2$ CDW devices. Even for devices showing high radiation tolerance, complex and expensive testing protocols are required for device qualification, which is not the case for these CDW devices [10,16,36,37].

In conclusion, we demonstrated that quasi-2D CDW devices have remarkable immunity to irradiation by high-energy protons. The *I-V* and LFN analysis show no significant changes up to at least the high fluence of $10^{14}$ H$^+$cm$^{-2}$, indicating the absence of both total-ionizing-dose and displacement-damage effects. The radiation immunity of the CDW devices is attributed to their two-terminal design, quasi-2D nature of the active channel, and the high concentration of the charge carriers. The results show that these devices are quite promising for the design of future radiation-hard electronics for critical systems required in the extremely high radiation environments associated with space applications, particle accelerators, nuclear reactors, and other extreme environments.

**METHODS**

**1*T*-TaS$_2$ crystal growth:** Millimeter-sized crystals of 1*T*-TaS$_2$ were grown via chemical vapor transport (CVT). Elemental Ta (Sigma-Aldrich), elemental S (J.T. Baker), and elemental I$_2$ (J.T. Baker) were added to a fused quartz ampoule. The ampoule was evacuated and backfilled three times with argon, with cooling to mitigate $I_2$ sublimation. The ampoule was flame-sealed and heated in a two-zone tube furnace at 10 °C min$^{-1}$ to 975 °C (hot zone) and 875 °C (cool zone). Then the ampoule was removed from the hot furnace and immediately quenched in a water–ice–NaCl bath. The structure and phase purity of the 1$T$-TaS$_2$ were verified with powder x-ray diffraction and HRTEM, and the correct stoichiometry was confirmed by energy dispersive spectroscopy.

**Device fabrication:** The devices were fabricated by the shadow mask method to avoid damage from chemical contamination typically associated with conventional lithographic lift-off processes. The shadow masks were fabricated using the double-side polished Si wafers with 3 µm thermally grown SiO$_2$ on both sides (Ultrasil Corp.; p-type; <100>). The details of the mask fabrication process have been reported elsewhere [38]. At the next step, the shadow masks were used to fabricate several devices by aligning the masks with the pre-selected layers of 1$T$-TaS$_2$ under an optical microscope. The aligned mask and device substrate were clamped together and placed in an electron beam evaporator (EBE) for contact deposition (10 nm Ti and 100 nm Au) through the mask openings. The completed devices were then transferred to another vacuum chamber for electrical characterization. The thickness and width of 1$T$-TaS$_2$ layers were determined using the atomic force microscopy (AFM) and scanning electron microscopy (SEM).

**Electronic noise measurements:** The noise spectra were measured using a dynamic signal analyzer (Stanford Research) and a semiconductor analyzer (Keysight). The measurements were conducted in the two-terminal device configuration. Since the contact resistance of the devices was low, the noise response was dominated by the contribution from the active device channel contribution. The extracted contact noise from the transmission line measurement structures was substantially below the thermal noise, which had no significant contribution to the channel noise. The dynamic signal analyzer measured the absolute voltage noise spectral density, $S_V$, of a parallel resistance network of a load resistor, $R_L$, and device under test, $R_D$. The normalized current noise spectral density, $S_I$, was calculated from the equation: $S^2=S_V\times[(R_L+R_D)/(R_L\times R_D)]^2/(G^2)$, where $G$ is the amplification of the low-noise amplifier.

**Proton irradiation**: Proton irradiation testing was conducted under vacuum at RT using the Vanderbilt Pelletron [24]. All devices were irradiated with both terminals grounded. The DC *I-V* measurements were conducted using a HP4156B semiconductor parameter analyzer during both the irradiation and noise measurement sequences. Forward and reverse sweeps were performed to observe the hysteresis window. One terminal was grounded during the *I-V* measurements, and the bias range for the other terminal was decided by the NC to IC CDW transition point. Details of the proton irradiation testing have been reported elsewhere in the context of other material systems [10,16].

**Acknowledgements**
The work at UCR and UGA was supported, in part, by the National Science Foundation (NSF) Emerging Frontiers of Research Initiative (EFRI) 2-DARE project: Novel Switching Phenomena in Atomic $MX_2$ Heterostructures for Multifunctional Applications (NSF EFRI-1433395). A.A.B. also acknowledges the UC - National Laboratory Collaborative Research and Training Program - University of California Research Initiatives (LFR-17-477237). Nanofabrication was performed in the Center for Nanoscale Science and Engineering (CNSE) Nanofabrication Facility at UC Riverside. The work at Vanderbilt University was supported, in part, by U.S. Air Force Office of Scientific Research (AFOSR) through the Hi-REV program. S. R. also acknowledges partial support from the Center for Terahertz Research and Applications co-financed with the European Regional Development Fund.

**Contributions**
A.A.B. conceived the idea, coordinated the project, contributed to experimental data analysis and led the manuscript preparation; A.G. fabricated devices and contributed to the experimental data analysis; E.X.Z. and S.E.Z. conducted proton irradiation testing and noise measurements; M.A.B. conducted material synthesis and characterization; T.T.S. supervised material synthesis and contributed to materials characterization; D.M.F. supervised proton irradiation testing and noise measurements, and contributed to data analysis; S.R. contributed to noise and radiation data analysis; E.A. conducted Raman and TEM material characterization. A.F performed simulation of proton irradiation. All authors contributed to manuscript preparation.

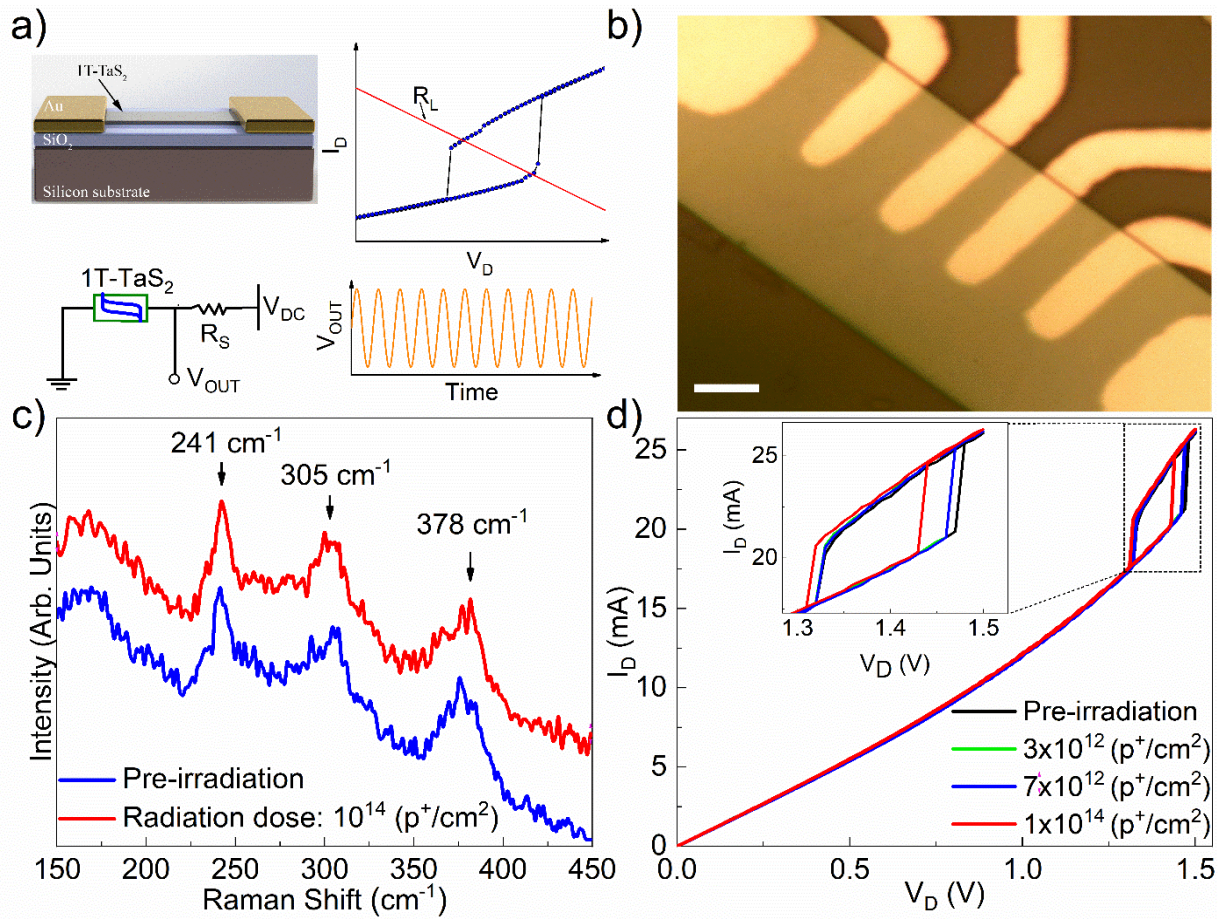

**Figure 1: Charge-density-wave device structure and characteristics.** (a) Schematic diagram of the two-terminal 1*T*-TaS$_2$ CDW device, its typical current-voltage characteristic, oscillator output and symbol representation (clockwise). (b) An optical microscopy image of fabricated 1*T*-TaS$_2$ device with Ti/Au (10 nm/100 nm) contacts. The scale bar is 3 µm. (c) Raman spectra of 1*T*-TaS$_2$ channel before (blue line) and after (red line) proton irradiation. No changes, which would indicate possible material damage, are observed. (d) Current-voltage characteristics of 1*T*-TaS$_2$ device after different proton fluences. The inset shows the hysteresis window, which is the working part of the *I-V* characteristic of the device. The changes in the current and voltage are negligible even after the maximum proton irradiation dose.

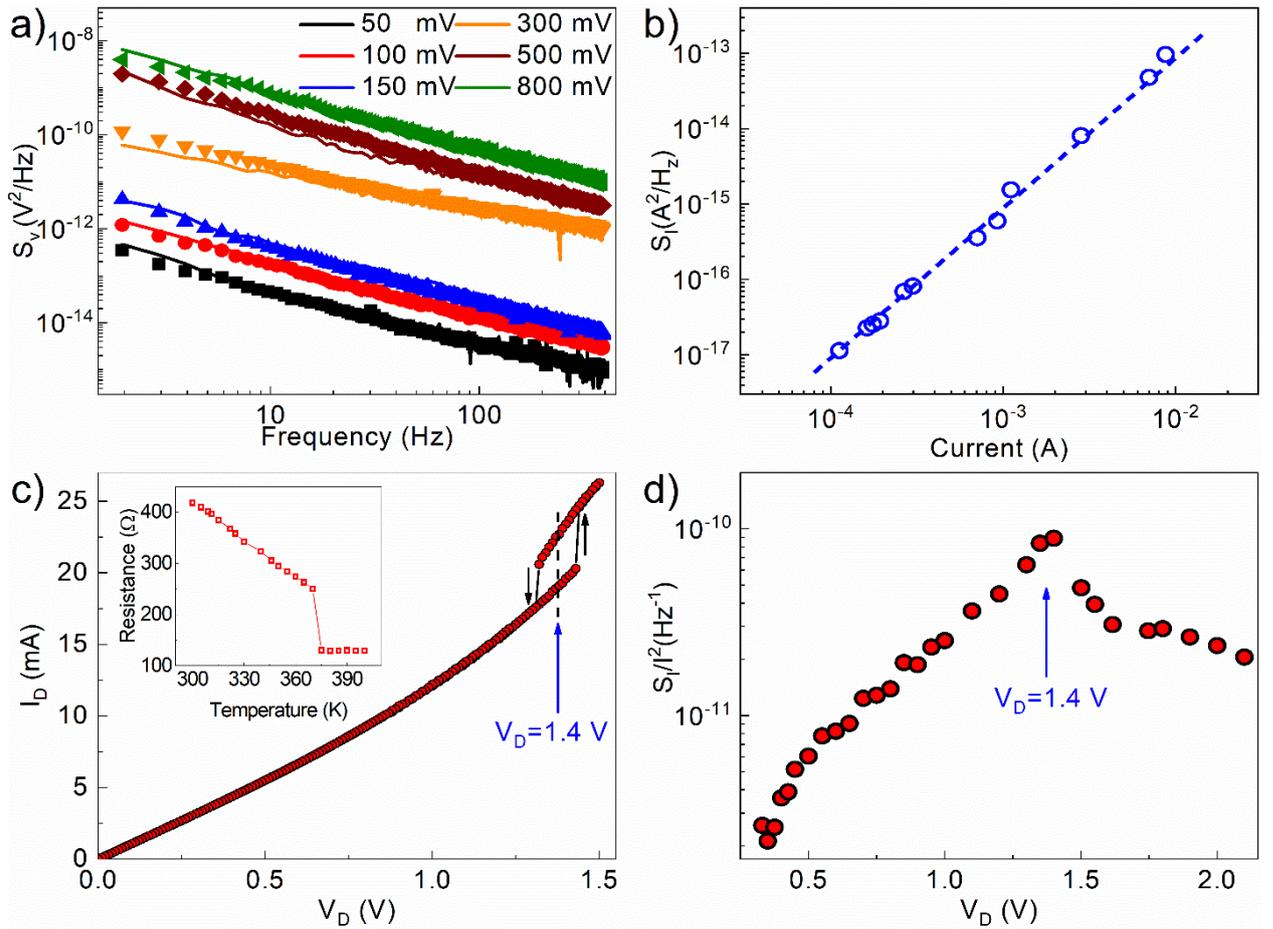

**Figure 2: Low-frequency noise and I-V characteristics of the charge-density-wave devices.** (a) The noise voltage power spectral density, $S_V$, as a function of frequency for a representative 1$T$-TaS$_2$ device before (solid lines) and after (symbols) proton irradiation. (b) Noise current power spectral density, $S_I$, as a function of the source-drain current, $I$, indicating the noise scales as $I^2$. Data are shown for the device after the maximum proton irradiation dose, at the frequency $f$=10 Hz. (c) Current-voltage characteristics of 1$T$-TaS$_2$ device after proton irradiation to $10^{14}$ P$^+$/cm$^2$, showing that the hysteresis window is preserved. The inset illustrates the temperature change of the resistance. (d) The normalized current noise spectral density, $S_I/I^2$, as a function of the source–drain bias. Note that the noise reaches its maximum level at a voltage of 1.4 V, which corresponds to the hysteresis window of the near-commensurate to incommensurate phase transition presented in the panel (c).